# The "rust" challenge: On the correlations between electronic structure, excited state dynamics and photoelectrochemical performance of hematite photoanodes for solar water splitting

Daniel A. Grave, Natav Yatom, David S. Ellis, Maytal Caspary Toroker*, and Avner Rothschild*

Department of Materials Science and Engineering, Technion – Israel Institute of Technology, Haifa, Israel
*Email: avner@mt.technion.ac.il, maytalc@technion.ac.il

**Abstract**

In recent years, hematite's potential as a photoanode material for solar hydrogen production has ignited a renewed interest in its physical and interfacial properties, which continues to be an active field of research. Research on hematite photoanodes provides new insights on the correlations between electronic structure, transport properties, excited state dynamics and charge transfer phenomena, and expands our knowledge on solar cell materials into correlated electron systems. This research news article presents a snapshot of selected theoretical and experimental developments linking the electronic structure to the photoelectrochemical performance, with particular focus on optoelectronic properties and charge carrier dynamics.

**Manuscript**

**Introduction**

Hydrogen extracted from water, with the help of sunlight, can be a renewable, non-polluting fuel.[1] In a photoelectrochemical cell, whose concept is illustrated in Figure 1a, light is absorbed in a photoanode material causing generation of electron-hole pairs. The holes move to the surface to drive the oxygen evolution reaction (OER) and the electrons transport through an external wire towards the cathode to drive the hydrogen evolution reaction (HER) as shown in Figure 1b. Since the seminal report on photoeletrochemical water splitting in 1972 with $TiO_2$ photoanodes,[2] numerous semiconducting absorber materials have been studied.[3–5] However, no material has yet satisfied all the requirements: stability under operating conditions, low cost, and high efficiency. Hematite (α-$Fe_2O_3$), one of the most widely studied materials as a photoanode for solar water splitting, has potential to meet these requirements. It possesses advantageous properties such as stability in alkaline solutions,[6] wide abundance, and visible light absorption.[7,8] The first study examining hematite as a photoanode material for solar water splitting dates back to 1976, showing that water photo-oxidation takes place for wavelengths longer than 400 nm.[9] While water photo-oxidation current densities (photocurrent in short) for hematite photoanodes have improved dramatically with the use of nanostructuring[10–13] or light trapping in ultrathin films,[14] they are still significantly lower than the theoretical limit based on the total amount of light absorption (12.6 mA/cm$^2$ for AM1.5G solar illumination).[14] The main reason for this has been attributed to significant charge carrier recombination, both in the bulk and at the surface.[15] Additionally, a large overpotential is required to drive the oxygen evolution reaction[16–18] which reduces the practicality of unassisted solar water splitting.[19] General aspects of hematite photoanodes[8,16,20] as well as specific topics such as simulations,[21,22] heterostructures,[23] kinetics,[24] surface states,[25] and underlayers and overlayers[26] have been previously discussed elsewhere. In this research news article, we highlight some of the recent experimental and theoretical work that links the electronic structure of hematite with its excited state dynamics and photoelectrochemical performance, and reflect on the related challenges towards making more efficient photoanodes. We focus on the processes outlined in the photoanode energy band diagram in Figure 1b: light absorption and generation of excess charge carriers, charge relaxation and transport, surface electrochemical reaction, and recombination both in the bulk and at the surface.

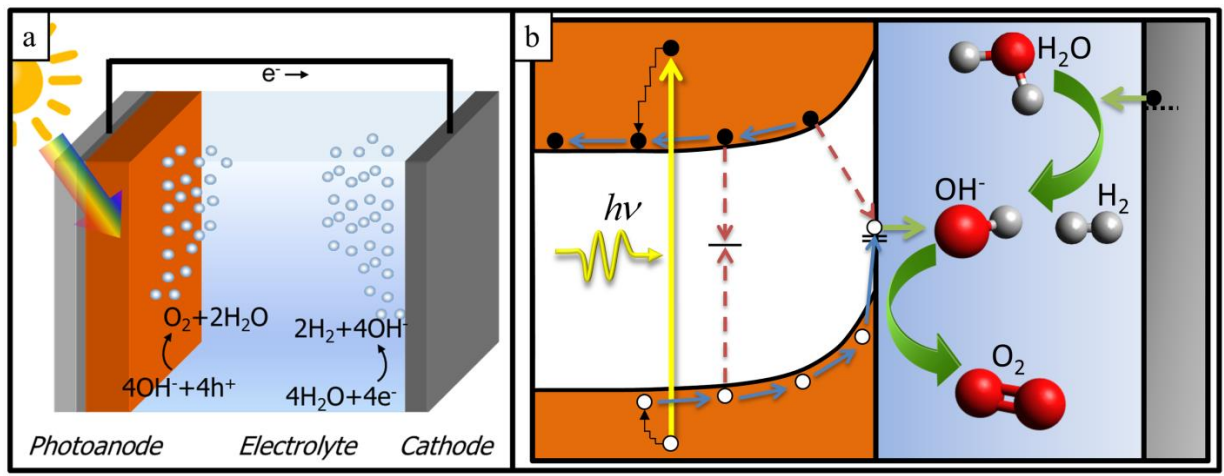

**Figure 1. Photoelectrochemical cell illustration and energy band diagram.** (a) Schematic diagram for the operation of a photoelectrochemical (PEC) cell: the photoanode (left) absorbs light, driving oxidation and reduction of water on the photoanode (left) and cathode (right), respectively. (b) Energy band diagram, describing electron (black circle)-hole (white circle) photogeneration (yellow arrows), thermal relaxation (stepped black arrows), transport (blue arrows), recombination (dashed red arrows), and reaction processes (green arrows). Red and gray spheres denote oxygen and hydrogen atoms, respectively. Holes transfer to the photoanode/water interface where they oxidize OH$^-$ adions through surface states (black horizontal lines), activating the OER, while electrons travel to the photoanode back contact, and through an external wire to the cathode, where they activate the HER.

**Electronic Structure**

Much is still unknown about hematite photoanodes, especially with regards to the relationship between the electronic structure, excited state dynamics and photoelectrochemical performance. The complexity of the electronic structure arises from the strong electron correlation in this material as a result of its crystal structure and composition. Hematite possesses a corundum crystal structure and may be visualized as a hexagonally close packed array of O atoms with Fe atoms in 2/3 of the octahedral sites leading to a crystal field splitting of Fe 3$d$ orbitals into the $t_{2g}$ and $e_g$ states.[27,28] Effectively, the iron atoms form basal planes perpendicular to the [0 0 1] direction, with the oxygen ligands lying between these planes, as shown in Figure 2a. There are five unpaired valence electrons in the Fe$^{3+}$ 3$d$ orbitals that are spatially localized and hence have a strong coulombic repulsion.[29–32] The resulting Fe-O bond is partially ionic and covalent in nature, as calculated by Bader charge analysis to have an effective charge of +1.8 on iron and -1.2 on oxygen using the conventional Density Functional Theory (DFT)+U method.[18,31,33] The "+U" factor has been introduced more than a few decades ago[34,35] to address the strong correlation of open shell transition metal $d$ orbitals, positioning hematite as a benchmark for correlated electronic structure calculations. The energy band structure of hematite has been partially resolved using state-of-the-art electronic structure theories, such as wave-function based methods, hybrid functionals, and many-body Green's function method with screened Coulomb potential $W$, known as the GW approximation

(GWA).[36–38] The *ab-initio* calculations were used to resolve measured spectra obtained by site-specific photoemission and inverse photoemission spectroscopy (PES/IPES), X-ray absorption spectroscopy (XAS), and Auger-electron spectroscopy.[37,39] The results show that the hematite valence band edge is dominated by O 2*p* hybridized with Fe 3*d* states, and the conduction band minimum is dominated by Fe 3*d* states, as shown in Figure 2b.

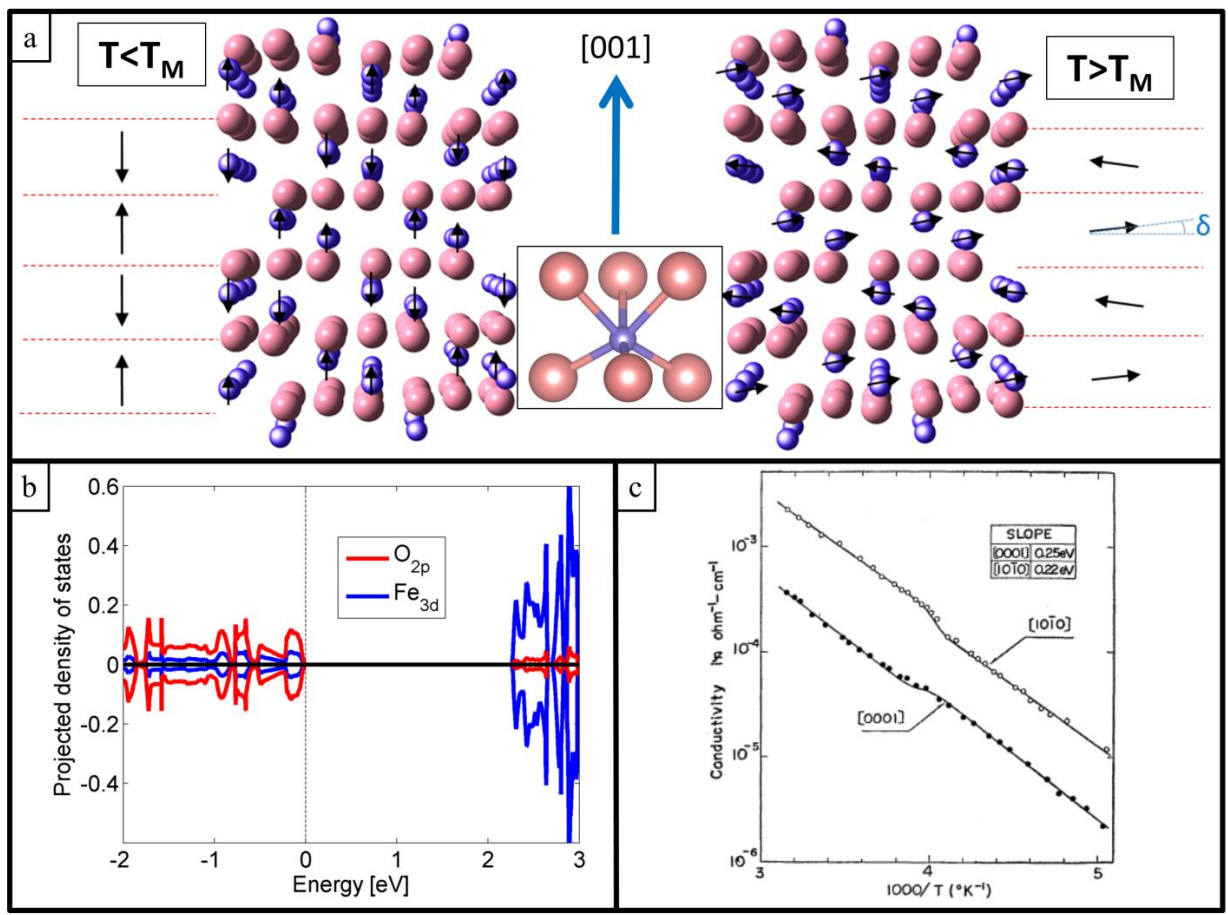

**Figure 2. Hematite magnetic and electronic structure.** (a) The crystal structure and spin ordering in hematite for T < $T_M$ (left) and T > $T_M$ (right). The iron atoms are represented by small blue spheres, and arrows pointing in their spin directions, and the large red spheres are oxygen atoms separating the Fe planes. Within each plane, the spin directions are the same, as indicated by the stacks of arrows on each side. For T > $T_M$, the spins cant out of the basal plane with an angle δ, as shown on the right hand side. The inset shows the octahedral coordination of the $Fe^{3+}$ ion. Adapted with permission from ref. [40]. Copyrighted (2017) by the American Physical Society. (b) DFT+U calculated projected density of states (PDOS) on O 2*p* (red) and Fe 3*d* (blue) in hematite. The PDOS is normalized to a maximum total density of one. Adapted with permission from ref. [41]. Copyright (2015) American Chemical Society. (c) Temperature-dependent electrical conductivity measured along the [001] and [100] directions of a hematite single crystal. Reproduced with permission from ref. [42]. Copyright by the Physical Society of Japan.

A consequence of the existing open shell Fe 3*d* orbitals in hematite is the appearance of spin ordering, which may have an impact on charge carrier dynamics as found for other systems where there is spin-orbit coupling and associated selection rules for charge transport.[43] The magnetic structure of hematite has been characterized by a number of experimental techniques.[44–47] Each $Fe^{3+}$ site has a spin of 5/2, and these spins are ordered throughout the crystal up to high temperatures (~ 950 K). The magnetic structure is depicted in Figure 2a. At low temperatures, the spins on each basal plane point perpendicular to the plane, along the [0 0 1] direction. Within each of these planes of iron atoms, the coupling is ferromagnetic (FM), with the spins of all of the atoms in the plane pointing in the same direction. Between planes, where the magnetic interaction is mediated by oxygen atoms, the coupling is antiferromagnetic (AF), with the spins alternating up and down. As the temperature is raised, the spins re-orientate so as to be nearly parallel to the basal plane, while still maintaining FM coupling within the plane, and mostly AF coupling (with regards to the in-plane spin direction) between planes. However, in this higher temperature state, a slight out-of-plane tilt of the spins is in the same direction for all basal planes, resulting in a "weak ferromagnetism". The transition between these two states is known as the Morin transition, which occurs in bulk, undoped hematite at $T_M \sim$ 265 K.[48] $T_M$ is sensitive to doping.[48–51] The magnetism in thin films has been recently begun to be characterized.[40,52,53] There have been experimental indications of an effect of the spin ordering state on electronic properties, as characterized by temperature-dependent electrical conductivity measurements[42,54] and shown in Figure 2c. However, most theoretical calculations have only considered the AF spin state. The weak FM spin-canted state arises from spin-orbit interaction. To date there has been only one calculation for hematite which has considered the weak FM state[55], and no experimental or theoretical studies have examined possible effects of magnetic state on the photoelectrochemical behavior of hematite photoanodes.

**Charge Transport**

Charge transport in hematite is often described using the small polaron model[56] where the movement of electronic charge carriers is strongly coupled to distortions of nearby atoms as they hop from site to site,[57–60] making the effective mass high and mobility low, as compared to conventional semiconductors.[55,61,62] An activated polaron hopping process is supported by temperature-dependent electrical conductivity measurements as shown in Figure 2c.[54,61,63,64] Recently, a change in the activation energy of electrical conductivity below the Morin transition was reported, further suggesting that charge transport is affected by the magnetic state.[54] Spin interactions as well as polaronic transport in hematite make the Hall effect anomalous,[54,61,65] which complicates using the standard method for measuring charge carrier concentration and mobility. Instead, most estimates of electron ($\mu_n$) and hole ($\mu_p$) mobilities have been derived using thermopower measurements or other techniques, with results usually in the $\mu_n \sim$ 0.01 cm$^2$/(Vs) range[14,63,66–68] and $\mu_p \sim$ 0.0001 cm$^2$/(Vs).[59,63] In single crystal hematite, there is a large anisotropy in electronic conduction (see Figure 2c), which has been reported to be from one to three orders of magnitude higher within the basal plane than in directions perpendicular to it.[42,55,58,69]

Utilizing the higher electronic conductivity within the basal plane has been suggested as a possible route of increasing the photocurrent by designing oriented nanostructures or films.[70–72] However, a recent study on epitaxial Sn-doped hematite thin film photoanodes with well-defined orientations observed that improvements in water photo-oxidation performance were not a result of improvement in hole transport from the bulk to the surface under illumination due to higher conductivity but rather due to orientation-dependent surface properties.[73] Doping is commonly used in order to improve both the photocurrent and onset potential for water photo-oxidation.[11,18,32,55,60,74–84] However, the role of doping on the photoelectrochemical performance of hematite photoanodes is not fully understood and has been attributed to a variety of factors including improvement in conductivity,[85] passivation of surface states and grain boundaries[86], shifting of band edge positions,[32,55,80] reduction in effective mass,[55] and distortion of the crystal structure[82–84] which facilitates hopping for both electrons and holes.[59]

**Light Absorption**

The ability to transport and extract photogenerated charge carriers under illumination is essential for the performance of hematite photoanodes. The incident photon conversion efficiency (IPCE) links the illumination provided by the solar spectrum with the generated photocurrent as a function of wavelength. A common trend observed in many IPCE measurements of hematite photoanodes is that the IPCE spectrum generally does not trace the absorbance spectrum, especially at higher wavelengths as shown in Figure 3a. This behavior is not typically seen in conventional solar cell materials such as silicon[87] or other photoanode materials such as ZnO[88] whose IPCE spectra more closely resemble an ideal square shape without the peak structures commonly observed in hematite photoanodes.[89] The mismatch between absorption and photocurrent was first noted by Kennedy and Frese and was attributed to optical excitations which did not contribute to the photocurrent.[90]

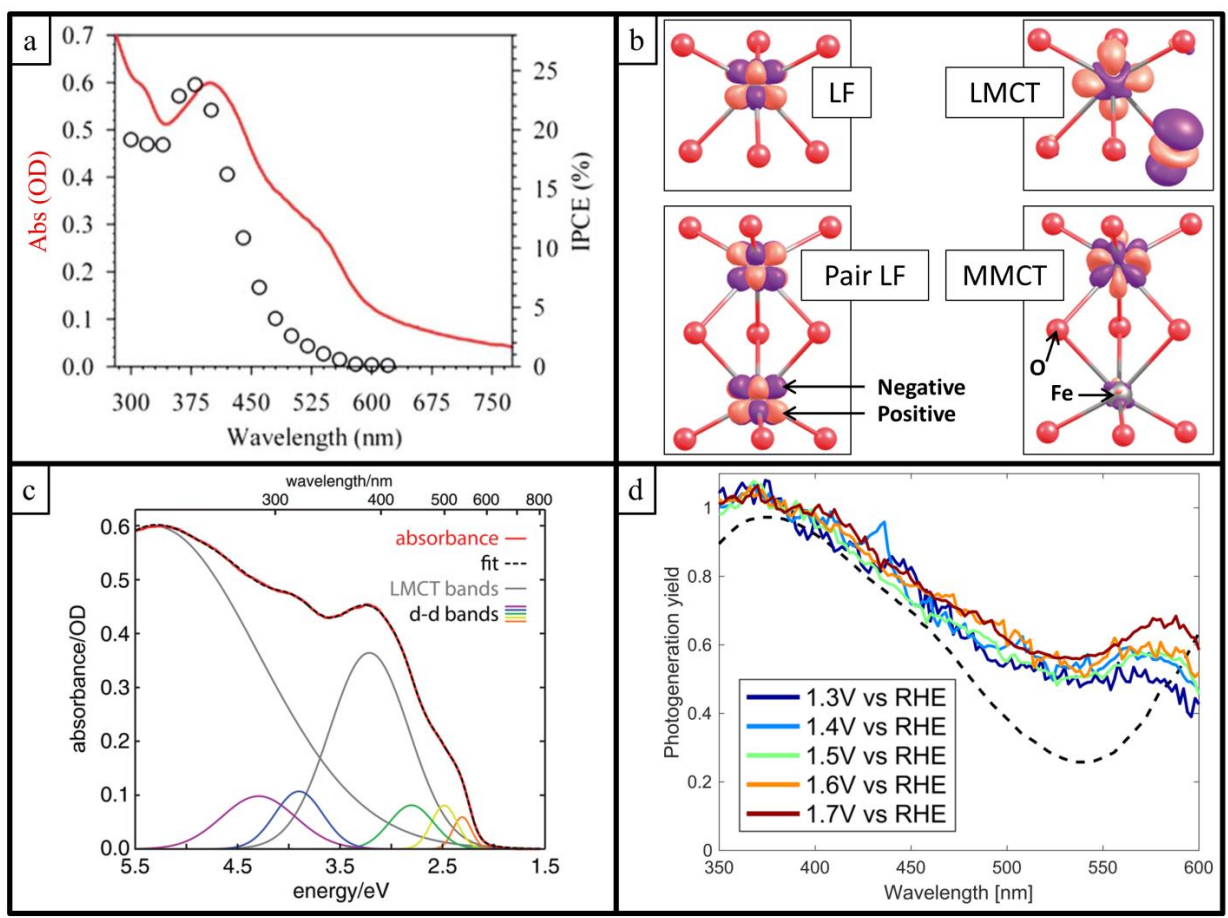

**Figure 3. Optically-induced electronic transitions and their impact on absorption and photocurrent action spectra.** (a) UV-Vis absorption (red line) and IPCE (black open circles) spectra of a 30 nm hematite thin film grown by atomic layer deposition. IPCE measurement conditions: 1.43 V *vs.* RHE, simulated AM 1.5 solar illumination, 1 M NaOH. Adapted from ref. [91] with permission of the Royal Society of Chemistry. (b) Calculated charge density difference iso-surfaces of four types of transitions in hematite: ligand field (LF), pair LF, ligand to metal charge transfer (LMCT), and metal to metal charge transfer (MMCT). Red and gray spheres represent oxygen and iron atoms, respectively. Iso-surfaces for positive and negative density differences are in pink and purple, respectively, which were obtained by subtracting the ground-state electron-density from the exited-state electron-density. Adapted with permission from ref. [29]. Copyright (2011) American Chemical Society. (c) The reflectance-corrected steady-state optical absorbance spectrum of a 20 nm thick undoped hematite film (red line). The spectrum was fit (dashed black line) to the sum of seven Gaussian bands, which are plotted individually. The absorption bands assigned to LMCT transitions are shown in gray, and those assigned to *d*–*d* transitions are shown in color. Reproduced from ref. [89] with permission of the Royal Society of Chemistry. (d) The photogeneration yield spectra extracted using spatial collection efficiency analysis of a 26 nm thick

heteroepitaxial 1% Ti-doped hematite film. Also shown in dashed black line is the spectrum calculated from ref. [89] considering only contribution of the LMCT bands to the photogenration. The color code in panel (d) represents the applied potential, as depicted in the legend. Adapted from ref. [87] with permission of Elsevier.

The absorption bands in hematite, originating from electronic transitions involving the $Fe^{3+}$ 3*d* orbitals, determine the nature of the excited charges. Four types of transitions have been considered to occur in hematite as depicted in Figure 3b: 1) Single ligand field (LF) transitions; 2) Pair LF transitions; 3) Metal to metal charge transfer (MMCT); and 4) Ligand to metal charge transfer (LMCT) transitions.[92] The single LF (also known as *d-d* or crystal field) transition involves a reorientation of an electron from one Fe 3*d* orbital to another on the same Fe site. The energy cost of this transition is the crystal field splitting energy. The pair LF transitions (sometimes referred to as double *d-d*, pair excitation, and double exciton) involve simultaneous LF transitions on adjacent Fe sites. The energy cost for this transition is expected to be roughly twice that of the single LF transition. In perfect octahedral coordination, the LF transitions are optically forbidden. However, the selection rules for the LF transitions can be relaxed by magnetic or vibronic coupling, or changes of crystal symmetry, so as to increase the probability of these transitions. The MMCT transition is when an electron transfers from one Fe site to another, as represented by the $2Fe^{3+} \rightarrow Fe^{2+} + Fe^{4+}$ disproportionation reaction. Sometimes in the literature, the MMCT transition has been referred to as a double *d-d* transition. However, here we distinguish that unlike the double LF transitions, the MMCT transition involves excitation across the Hubbard gap, creating an electron-hole pair on separate Fe sites. Lastly, the LMCT transition given by $Fe^{3+} + O^{2-} \rightarrow Fe^{2+} + O^{-}$ involves a transfer of an electron from an O 2*p* orbital to an Fe 3*d* orbital, creating an electron-hole pair on the Fe and O sites.

The roles of these various excitations in the absorption spectrum and photocurrent of hematite are still under debate. Over the years, numerous assignments have been given to features in the optical spectra on the basis of the aforementioned optical transitions, without a clear consensus.[29,93–96] Kennedy and Frese proposed that LMCT transitions yielding $O^-$ holes contributed to photocurrent, while the MMCT transition yielding $Fe^{4+}$ holes did not. As early as 1963, it was suggested that the energetic requirement for the MMCT transition was prohibitively large.[97] To this day, the prevalent view is that the LMCT energy is lower than the MMCT energy, classifying hematite as a "charge transfer insulator" as opposed to a "Mott-Hubbard insulator".[22] As for the other transitions, the onsite single and pair LF transitions do not directly create electron-hole pairs, although it has been suggested that a superexchange or hopping mechanism can result in the formation of the $Fe^{2+}$ - $Fe^{4+}$ electron-hole pair after excitation.[98] The excited states generated by the LF and pair LF transitions are considered to be highly localized compared to the holes generated by the LMCT transition. While it is generally accepted that the LMCT transitions are higher in energy than the single LF transitions, their energies have not been ruled out to be close or overlapping. Certainly, pair LF transition energies could overlap with the LMCT transition energies. The absorption edge is generally considered to be due to the LF transition,[99] though some reports assess that it possesses LMCT character.[94]

The onset of the LMCT transition is not clear from the optical absorption spectra nor manifested as a sharp increase in the IPCE spectra, as depicted in Figure 3a. This can possibly be explained by more recent assignments where the absorption spectrum was described by two LMCT bands which span the optical spectrum from the deep UV to the NIR, along with multiple LF bands which also span the whole spectrum, dominating at higher wavelengths as shown in Figure 3c.[89] This hypothesis was supported by reproducing experimental IPCE spectra (Figure 3a)[91] through calculation using only LMCT excitations as those that produce photocurrent.[89] In another study, the photogeneration yield $\xi(\lambda)$, defined as the probability of absorbed photons to generate mobile charge carriers, was recently extracted for a heteroepitaxial Ti-doped hematite photoanode[100] using empirical spatial collection efficiency analysis as shown in Figure 3d.[87] Good qualitative agreement was found between the extracted $\xi(\lambda)$ spectrum and the predicted one based on the band assignments in Figure 3c, assuming that only LMCT excitations contribute to the photocurrent. The discrepancy at high wavelengths in Figure 3d was explained by Ti-doping which has been suggested to reduce losses associated with the LF transitions.[101] Another demonstration of the effect of multiple types of excitations on hematite photoelectrochemical properties has been shown in a recent study of a solid 1 μm thick Ti-doped hematite film.[102] Back illumination yielded higher photocurrent for 450 nm excitation than 530 nm excitation despite the light being absorbed further away from the front surface. This shows that the photocurrent is strongly dependent on the optical excitation wavelength, yielding both localized and delocalized electronic transitions, and not only on the proximity of absorption to the front surface. Additionally, photogenerated charge carriers were shown to be collected from depths of as much as 700 nm, implying that the collection length for at least some photogenerated holes must be significantly longer than the commonly cited estimates of 2-4 nm based on fitting photoelectrochemical measurements to the Gartner model.[90,103] However, no other independent verification of the hole diffusion length exists. Accurate measurement of hole mobility and lifetime would give a better estimation of hole diffusion length.

**Charge Carrier Dynamics**

Insight into the charge carrier dynamics in hematite was obtained from transient measurements in the excited state immediately after illumination (pump probe measurements). Early transient absorption spectroscopy (TAS) measurements on epitaxial thin films and bulk single crystals attributed a fast 300 fs time constant to relaxation of hot electrons to the conduction band edge followed by recombination with holes and trapping in mid-gap states within ~3 ps.[104] The resulting trap states were suggested to live for hundreds of picoseconds or longer. The recent emergence of 4D electron energy loss spectroscopy (EELS) has allowed temporal analysis of the change in valence state of iron induced by an ultrafast laser pulse of visible light.[98] A change in width of the Fe *L* edge peak under 519 nm illumination was interpreted as a signature of *d-d* transitions that dominate the response at this wavelength. The signal was found to decay within ~3 ps. More recently, another model based on XUV spectroscopy at the Fe *M* edge and charge transfer multiplet calculations was suggested wherein small polaron trapping is assumed responsible for ultra-fast carrier localization after LMCT excitation.[105] Wavelength dependence of hematite IPCE in relation to absorption was explained by higher energy excitation giving rise to longer polaronic lifetime and increased hopping radius. Both ref. [105] and [98] measured similar ultrafast dynamics including a sharp ~150 fs peak and ~3 ps decay, but with fundamentally different interpretations. While the consistency of empirical findings using different techniques is encouraging, a fully consistent picture of the charge carrier dynamics in hematite has yet to be established. Another commonality in most transient studies is the observation of multiple decay processes, with some measurements spanning vastly different timescales from hundreds of fs to μs in the absence of applied electrical bias (potential).[91] The nature of these processes is still under discussion without established consensus on the assignment of the measured time constants. In summary, the ultrafast processes (~100 – 300 fs) have been attributed to initial polaron formation[105] or hot electron relaxation.[104] Fast (~few ps) processes have been attributed either to electron-hole pair recombination,[104] or to full decay of the initial charge transfer state to the polaronic state,[105] as well as trapping in surface or mid-gap bulk states.[104] The long lifetime processes (from tens of ps to μs) have also been attributed to charge carrier recombination,[91,106] trapping,[104] and thermal expansion and cooling effects.[89] The recently observed long collection length[102] in thick film Ti-doped hematite photoanodes may support assignment of mid-to-long decay processes observed in transient spectroscopy measurements to mobile charge carriers.

**Surface Reaction**

Beyond the charge carrier dynamics in the bulk, processes which govern interfacial charge transfer at the surface have been the focus of much study.[107,108] New techniques have offered insight into the electron dynamics of these processes. Surface sensitive reflective XUV spectroscopy of hematite showed that in the absence of an applied potential, ultrafast processes attributed to localization and small polaron formation occur on longer time scales at the surface than it does in the bulk.[109] Additionally, it was found that this ultrafast trapping had similar timescales for both polycrystalline and monocrystalline samples, suggesting that crystallographic defects such as grain boundaries do not affect carrier localization at the surface. Moreover, *operando* spectroscopic measurements demonstrate that the electrochemical conditions for water photo-oxidation dramatically change the surface dynamics. Two different electronic transitions were identified by *operando* soft X-ray absorption at the O *K*-edge under light and dark conditions, both of which were found to contribute to water oxidation.[110]

Durrant and co-workers also observed two electronic transitions by *operando* TAS measurements which show that charge carrier lifetimes extend to seconds for anodic potentials above the photocurrent onset potential.[15,111] The two main features in the TAS exhibited a remarkable correlation with the photocurrent. At potential well below the photocurrent onset, there is a narrow positive feature (increased absorption) at 580 nm, near the band edge. As the potential increases, this sharp feature changes sign and becomes increasingly negative, while simultaneously a positive, broad sub-bandgap continuum feature (peaked at 650 nm) emerges. Figure 4a shows that the magnitude of this latter sub-bandgap feature directly tracks the photocurrent vs. potential curve, suggestive of a surface state directly involved with water photo-oxidation.[111,112] This behavior was found to be independent of photoanode morphologies and Si-doping levels. The sharp 580 nm feature continuously increases with potential even before the onset of photocurrent, as shown in Figure 4a, and was attributed to trapping states.[111,112] We note that the evolution of the 580 and 650 nm peak amplitudes in Figure 4a closely resembles the photocurrent vs. potential curves with and without hole scavenger (such as $H_2O_2$),[113] suggesting involvement of these transitions with surface recombination and water photo-oxidation, respectively. The 580 nm feature was further investigated by other techniques. Interestingly, absorption difference spectra obtained by light- and potential-modulated absorption spectroscopy (LMAS/PMAS)[114,115] and steady state potential-dependent absorption difference,[112,116,117] were found to be similar as shown in Figure 4b with a peak at 580 nm. The 580 nm peak was found sensitive to overlayers, suggesting that it relates to the surface.[117] Despite all of these experimental fingerprints, the chemical identity of the surface states as well as their role in either water photo-oxidation or surface recombination is under debate.[111,114,115,117–119]

Recently, DFT+U calculations identified a possible origin of the 580 nm absorption peak as $Fe^{+4}O$ intermediate species of the oxygen evolution reaction (OER), shown in Figure 4c.[25,41] This assignment was later confirmed by *operando* infrared spectroscopy.[120] Further insight into the surface electronic states responsible for the spectra in Figure 4b was achieved from DFT+U calculated absorption difference spectra.[31] A good match was found between the calculated spectra for the $Fe^{+4}O$ surface species and the experimental spectra previously discussed as shown in Figure 4b. The calculations reveal that the measured spectral response originates from two kinds of transitions that involve surface states localized on the Fe$^{+4}$O intermediate. Referring to the density of states graph in Figure 4d, these transitions are: (1) Transitions from valence band electrons to hybridized O 2$p$ – Fe 3$d$ unoccupied mid-gap states located 0.5 eV above the valence band edge (see Figure 4d panel 2) allowing absorption far into the IR ; (2) Transition from occupied O 2$p$ surface states that overlap in energy with the valence band edge (Figure 4d panel 1) to Fe$^{+4}$ unoccupied surface states that overlap in energy with the conduction band edge (Figure 4d panel 3) allowing absorption around the bandgap. These two transitions may correlate to the IR continuum and 580 nm peak observed by TAS, respectively.[111]

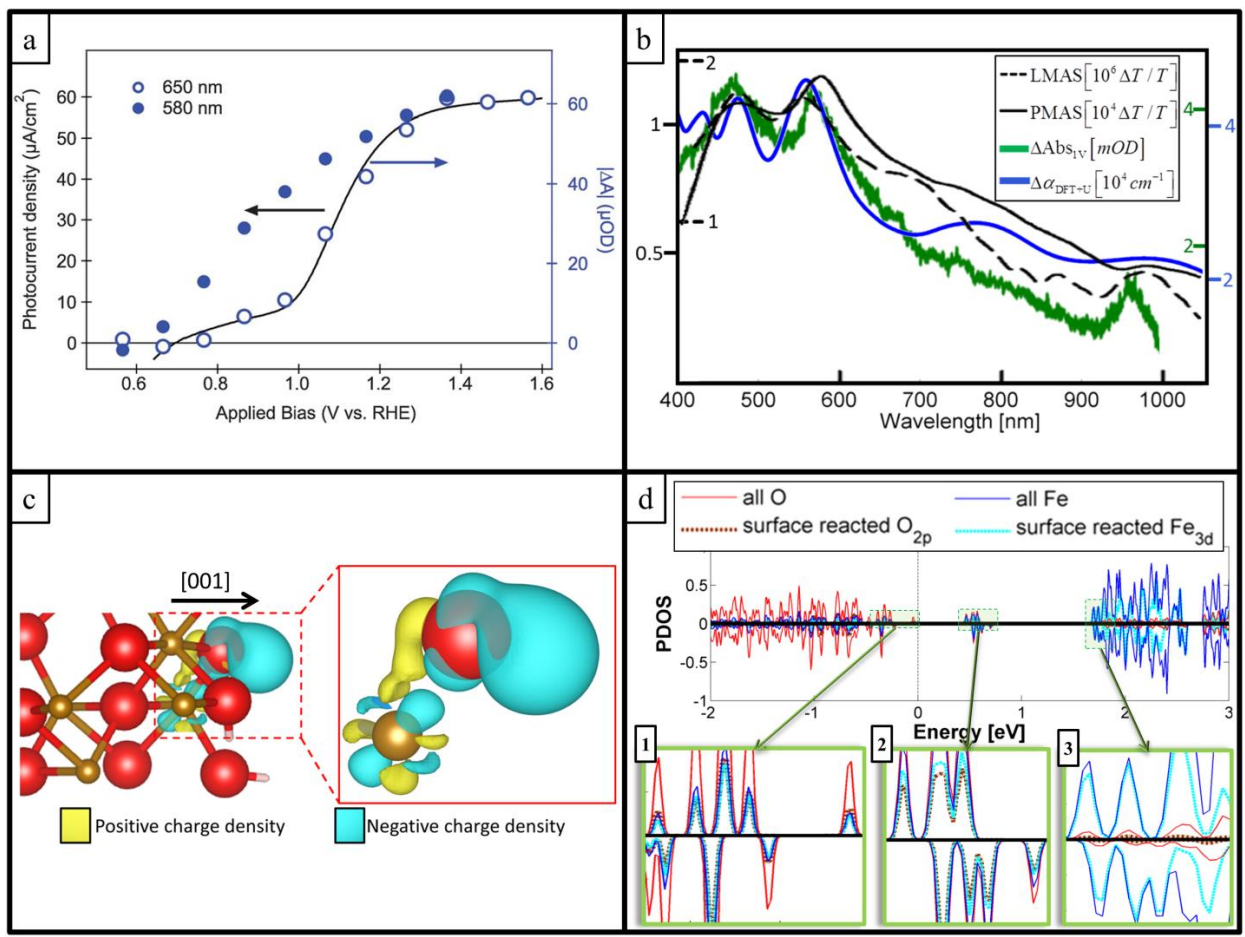

**Figure 4. Correlation of surface electronic structure with water photo-oxidation reaction.** (a) Correlation between the photocurrent (black curve) and absolute transient absorption amplitude (blue circles) as a function of applied potential. Open and closed circles correspond to the transient absorption difference at 650 and 580 nm, respectively, probed 100 ms following laser pulse excitation. Reproduced from ref. [111]. Published by the Royal Society of Chemistry. (b) DFT+U calculated absorption coefficient difference ($\Delta\alpha[cm^{-1}]$) between *O intermediate and the *OH neutral surface (blue),[31] overlaid on normalized transmission difference ($\Delta T/T$) spectra measured by PMAS and LMAS (solid and dashed black lines black, respectively)[114] and the absorption difference ($\Delta Abs[mOD]$) spectrum obtained by subtraction of the measured absorption spectrum at 1 V vs RHE from the spectrum at 2.2 V vs RHE (green).[117] All the spectra are normalized to their maximal value. Adapted from refs. [31,114,117]. (c) Charge density difference showing a surface state on pure hematite (001) surface.[41] Calculated by subtracting the electron density of the *OH intermediate from the *O intermediate at the fixed ionic positions of the latter. Red, gold, and white spheres denote O, Fe, and H atoms, respectively. Negative and positive charge density iso-surfaces (±0.01 e/Bohr$^3$) are in blue and yellow, respectively, where negative is the absence of electrons. Created with VESTA visualizing software.[121] Reproduced with permission from ref. [41]. Copyrighted by the American Physical Society. (d) Projected density of states (PDOS) with *O oxidation

intermediate on hematite (001) surface. O, Fe, and H total DOS is in red, blue, and green solid lines, respectively. Boxes zoom in on the surface states for the *O intermediate, with the 2*p* orbital of O atom at the active site, and the 3*d* orbital of the neighboring Fe atom in red and blue broken lines, respectively. The DOS is normalized to a maximum total density of one. Adapted with permission from ref. [41]. Copyright (2015) American Chemical Society.

**Overlayers**

Surface modifications have been used to reduce the overpotential and explore the effect of different surface states on water photo-oxidation. Annealing treatments and overlayers were found to give rise to cathodic shift of the photocurrent onset potential by few hundred mV.[122–127] The distinction between the effect of the overlayer to suppress surface recombination or to catalyze the OER has been studied by hole scavenger measurements,[122,128] intensity modulated photocurrent spectroscopy (IMPS),[114,128–130] TAS,[116] and dual working-electrode (DWE) techniques.[131,132] One of the most prominent examples for the effect of overlayers on photoelectrochemical performance is decrease of 0.4 V in the overpotential (OP) with significant increase of the photocurrent upon $Ni_{1-x}Fe_xO_y$ deposition.[127,128,131,133] The improvements in performance upon $Ni_{1-x}Fe_xO_y$ deposition have been attributed to surface passivation, enhanced catalysis and/or hole collection.[128,131,133] Deposition of $Al_2O_3$,[124,126,134] and $Ga_2O_3$,[124] thin films resulted in an OP decrease of 0.1 V and 0.2 V, respectively, which were attributed to passivation of hole trapping surface states. Deposition of an $Al_2O_3$ overlayer enabled measurable photoluminescence, that was not observed in bare hematite thin films, attributed to removal of surface traps.[126] Further understanding of the linkage between overlayers and photoelectrochemical performance was obtained by DFT+U calculations of water photo-oxidation on $Al_2O_3$[135] and $Ga_2O_3$[136,137] coated hematite with full or partial coverage of the overlayer. While calculations found that full coverage by $Al_2O_3$ prohibits water photo-oxidation,[135] partial coverage was calculated to reduce the OP by 0.13 V,[135] in excellent agreement with measurements.[124,126,134] The reduction in OP on the uncovered hematite sites for non-catalytic overlayers such as $Al_2O_3$ was attributed to an increase of surface band banding, which may assist the hole transport to the surface.[135] For the $Ga_2O_3$ overlayers, calculation showed that the valence band edge decreased resulting in higher OP.[136] However, experimental results[124] observed a decrease in OP possibly as a result of enhanced performance through other mechanisms, such as lower effective mass of $Ga_2O_3$[62] and passivation of surface states.[137]

**Conclusions and Future Outlook**

The correlated nature of hematite affects charge carrier generation and recombination, charge transport, and interfacial charge transfer, which have been shown to be generally complex, with multiple processes, timescales, or steps. Reports using different techniques to study these processes have often yielded similar observations of energies and timescales. While much progress has been made in the analysis of these observations, consensus of the specific interpretations and assignments to different features have yet to be established. Establishing a clear picture of the charge carrier dynamics will require multiple experimental techniques to bridge different timescales. Links should be established between ultrafast processes investigated by pump-probe methods such as TAS to slower processes occurring under longer timescales, such as those investigated by impedance spectroscopy and the related "distribution of relaxation times" analysis.[114,129,130] Moreover, it is important to link the optical and electrical information gained from these analytical methods with chemical information from other spectroscopic methods. For instance, X-ray photoemission is a well-established spectroscopy tool to identify chemical species, charges, and potentials, for which new developments will allow for ultrafast time-dependent, *operando* analysis, and/or nm-resolution depth profiling investigations.[138–141] Surface-sensitive *operando* infrared spectroscopy can also identify surface intermediates of the water oxidation reaction.[120] New in *operando* methods with high spatial resolution, such as a recently developed potential-sensing atomic force microscopy technique,[132] can yield valuable insight into photoanode operation. The large variety of techniques can each provide complementary information on one or more of the processes involved in photoelectrochemical water splitting: light absorption and carrier generation, charge transport, and surface electrochemical reaction.

One of the major issues related to the light absorption process are the ineffective optical transitions, which do not yield mobile charge carriers that contribute to the photocurrent. Future work would benefit from clearer understanding of what governs the wavelength dependent response of hematite photoanodes, focusing not only on the proximity of the absorption close to the surface, but also on the proportion of effective optical excitations which contribute towards generation of mobile charge carriers vs. ineffective excitations to localized states. One promising direction includes the use of doping to improve the IPCE in spectral regions where LF transitions dominate the optical response.[101] It has been suggested that using strain to shift the optical absorption spectrum towards the UV where the LMCT transitions dominate may yield more mobile charge carriers.[89] Likewise, it has been shown that by shrinking nanoparticle sizes, the structure can be manipulated to suppress the LF transitions, at the cost of increasing the band gap.[94]

Many questions remain open about the role of material properties on charge transport in hematite. The possible role of the spin ordered state on hematite photoelectrochemical behavior has largely been overlooked and should be investigated. Strain[52] and doping[40,53] in thin films cause dramatic changes in magnetic state. Furthermore, a charge density mapping experiment suggested that the Morin transition also corresponded to a transition from a charge transfer to Mott-Hubbard insulating state.[142] The role of crystallinity, orientation, and grain boundaries on the charge carrier dynamics, surface properties, and photoelectrochemical performance is not clear. TEM studies have shown the detrimental effect of high-angle grain boundaries on charge transport and photoelectrochemical performance of nanostructured hematite photoanodes.[10] Systematic studies using high quality epitaxial thin films can relate the electronic structure and excited state dynamics to the effects of crystalline structure and chemical composition (e.g., doping) without other spurious contributions that arise from grain boundaries. Furthermore, it is possible to grow heteroepitaxial films deposited in different orientations thereby exposing different facets at the surface.

Understanding of the surface water oxidation reaction has advanced significantly through DFT+U calculations.[60,62,81,82,143,144] In particular, a successful comparison has correlated between the electronic structure of surface reaction intermediates and the experimentally measured absorption spectra.[31,41] Furthermore, study of overlayers on hematite has also yielded valuable insight into modifying the interfacial electronic structure for enhanced water photo-oxidation.[135–137,145] However, these studies focus on calculating thermodynamics properties such as the free energies required for the intermediate reaction steps. Extending these studies to calculating chemical kinetics as well as associated charge carrier dynamics at the surface will facilitate enhanced understanding of catalysis, charge transfer and recombination processes.

While much work is focused on overlayers and co-catalysts, modification of the back contact or insertion of underlayers has also yielded significant performance enhancements.[146] This demonstrates that surface, interface and bulk processes are intertwined. Asymmetry, which is critical for charge separation and extraction in solar cell materials,[147] can be enhanced by selecting proper back contacts[148] and front contacts (overlayers). Empirical spatial collection efficiency analysis enables study of asymmetry in *operando* conditions and provides additional insight into productive photogeneration yield.[87]

We anticipate that a combination of experimental and theoretical efforts that include well-controlled fabrication and time-dependent kinetic and dynamical calculations that account for transition states and excited states will shed light on the fundamental relation between electronic structure, charge carrier dynamics, and photo-conversion efficiency and provide valuable information that will hopefully direct future efforts to rational design of high performance photoanodes.

**Acknowledgement**

This research has received funding from the European Research Council under the European Union's Seventh Framework programme (FP/200702013) / ERC (grant agreement n. 617516), from the Nancy and Stephen Grand Technion Energy Program (GTEP), from the I-CORE Program of the Planning and Budgeting Committee, and The Israel Science Foundation (Grant No. 152/11), and from the Ministry of Science and Technology of Israel. D.A.G. acknowledges support by Marie-Sklodowska-Curie Individual Fellowship No. 659491. D. S. Ellis acknowledges support from The Center for Absorption in Science at the Ministry of Aliyah and Immigrant Absorption in Israel. N. Y. acknowledges an excellence scholarship by the Zeff Award.